\documentclass[12pt]{article} \usepackage{amsmath}
\usepackage{amsfonts} \usepackage{amssymb} \usepackage{latexsym}
\usepackage{graphicx} \usepackage[english]{babel} \input epsf.sty

\begin{document}

\title{
{\normalsize \hskip4.2in USTC-ICTS-07-07} \\{The Measure for the Multiverse and the Probability for
  Inflation }}

\vspace{3mm} \author{{Miao Li$^{1,2}$, Yi Wang$^{2,1}$}\\
{\small $^{1}$ The Interdisciplinary Center for Theoretical Study}\\
{\small of China (USTC), Hefei, Anhui 230027, P.R.China}\\ {\small
$^{2}$ Institute of Theoretical Physics, Academia Sinica, Beijing
100080, P.R.China}} \date{}
\maketitle

\begin{abstract}
We investigate the measure problem in the framework of
inflationary cosmology.
The measure of the history space is constructed and applied to inflation
models. Using this measure,
it is shown that the probability for the
generalized single field slow roll inflation to last for $N$ e-folds is suppressed by a
factor $\exp(-3N)$, and the probability for the
generalized $n$-field slow roll inflation is suppressed by a much
larger factor $\exp(-3nN)$. Some non-inflationary models such as the
cyclic model do not suffer from this difficulty.
\end{abstract}

\newpage
\section{Introduction}

It is claimed that our vacuum is one of the $10^{500}$ possible
meta-stable vacua in the string theory landscape \cite{landscape}.
If this is true, then the physical parameters labeling which vacuum
we are living in can not be calculated from the first principle.
Theoretically, these parameters may only be explained by some
anthropic reasoning \cite{W89}, or by pure chance.

From the cosmological point of view, in the framework of eternal
inflation \cite{eternal}, the vast landscape of vacua is not only a
logic possibility but also the reality. If we demand that our
observable universe is not too special in the multiverse, in
principle, we can make predictions in the multiverse by calculating
the probability of the corresponding universe history.

A serious problem arises at this point.  A measure of the history space is essentially needed in order to
 compare different histories of the universe. But in general
relativity, it is not straightforward to construct such a measure. It is
because there is no preferred space slicing and time notation in
general relativity, and singularities commonly arise in the cosmic
solutions. Even in the much simplified Friedmann-Robertson-Walker
universe, the measure problem is not easy to solve. The construction of a measure of the history space is
considered as one of the central problems in cosmology. Attempts for
this problem can be found in \cite{GT06, attempts}.

To analyze this problem in detail, let us construct the history space of the
universe and discuss the measure.
In general relativity, all the trajectories in the phase space should
lie on the hypersurface $\mathcal{H}^{-1}(0)$ due to the Hamiltonian
constraint $\mathcal{H}=0$. Now we want to consider the history space,
where a trajectory is represented by a single point. So we have to
identify the points in $\mathcal{H}^{-1}(0)$ which can be linked by
the time evolution. Then, the history space, or the multiverse, takes
the form
\begin{equation}
  \label{eq:M} M=\mathcal{H}^{-1}(0)/\Bbb{R}.
\end{equation}

The next step is to construct a measure on the history space. To make
sense and to be natural in physics, a measure of the history space
should satisfy three conditions \cite{GT06, GHS86}: (i) It should be
positive. (ii) It should depend only on the intrinsic dynamics and
neither on any choice of time slicing nor on the choice of dependent
variables. (iii) It should respect all the symmetries of the space of
solutions.

A measure of the history space satisfying these three requirements can
be constructed from the phase space symplectic form \cite{GT06, GHS86,
  HC83}. The symplectic form of the phase space $\omega$ can be written
in terms of the canonical coordinates and momenta as
\begin{equation}
  \label{eq:symplectic} \omega=\sum_{i=1}^m dp_i \wedge dq^i.
\end{equation}
where $m$ is the number of canonical coordinates.

If we choose $p_m=\mathcal{H}$, then from the Hamilton's equations,
$q^m=t$ is the time coordinate. And the symplectic form
(\ref{eq:symplectic}) can be written as
\begin{equation}
  \label{eq:symplectic2} \omega=\sum_{i=1}^{m-1} dp_i \wedge
dq^i+d\mathcal{H}\wedge dt.
\end{equation}

The Hamiltonian constraint $\mathcal{H}=0$ naturally yields a
two-form transverse to the time evolution. This is the two-form in
the history space,
\begin{equation}
  \label{eq:wc} \omega_C\equiv
\omega|_{\mathcal{H}=0}=\sum_{i=1}^{m-1} dp_i \wedge dq^i.
\end{equation}

The measure of the history space can be constructed by raising
$\omega_C$ to the $(m-1)$th power,
\begin{equation}
  \label{eq:WM} \Omega_M\equiv
\frac{(-1)^{(m-1)(m-2)/2}}{(m-1)!}\omega_C^{m-1}.
\end{equation}

Note that $\Omega_M$ is an exact form. It can be globally written as
$\Omega_M\sim dA$ with
\begin{equation}
  \label{eq:A} A\equiv p_1 dq^1 \wedge \sum_{i=2}^{m-1} dp_i \wedge
dq^i.
\end{equation}

This measure of the history space can be applied to the inflationary
cosmology in determining the probability of inflation. At first, it
was believed that the canonical measure favors inflation
\cite{GHS86}. But soon it is realized that both inflationary and
non-inflationary history have infinite measure \cite{HP88}. So the
measure problem in cosmology remained unsolved.

Recently, Gibbons and Turok \cite{GT06} suggested a solution to this
measure problem. They noticed that a universe with a very small
spacial curvature at the present time can not be distinguished from a flat
one. So physically, it makes sense to cut off the history space by
identifying a universe with a very small spacial curvature with a flat
universe. As was shown in \cite{GT06}, the measure for some quantities, like the
spacial curvature, is cutoff dependent, and dominated by the
cutoff. While the measure for some other quantities, for example, the
e-folding number of inflation, is cutoff independent. So by applying
this cutoff, the question whether a N e-folds' inflation is natural
can be well
defined, and investigated explicitly. It is shown that the
history space volume for slow roll inflation is suppressed by a factor of
$\exp(-3N)$, where $N$ is the e-folding number.

The work \cite{GT06} concentrates on a single field minimal coupled
inflation model. There is a vast variety of inflation models in addition
to a single inflaton model, thus it is interesting to ask how other models
weigh in this measure.
Some of the inflation models involve a modified Lagrangian density other
than the minimal one, some involve multi-fields and some  modify
the Einstein gravity. We want to know whether
these inflation models are also suppressed for a large e-folding number. An investigation
of these models is the main task of this paper.

This paper is organized as follows. In Section 2, we review the
approach by Gibbons and Turok \cite{GT06} for gravity minimally
coupled to a  scalar field. It is shown that the inflation
probability can be calculated directly as a function of $N$. In Section 3, we discuss the
measure for the scalar field with a more general Lagrangian. We find
that in this generalized case, the measure for the slow roll
inflationary history is suppressed by exactly the same factor
$\exp(-3N)$. In Section 4, we consider the multi-field inflation. It
can be shown that with the assumption of slow roll for the Hubble
constant, the measure is a lot more suppressed by
the exponential factor $\exp(-3nN)$, where $n$ is the number
of inflaton fields. So it seems much more unnatural for multi-field
inflation to happen. In Section 5, we investigate the generalized
Lagrangian for multi-field inflation. We find that the generalization
of the Lagrangian can not solve the measure problem raised in Section
4. Finally, we summarize the paper in the last section.

\section{Single Field Inflation Models}

In this section, we consider a single scalar field minimally coupled
with gravity with the action
\begin{equation}
  \label{eq:S2}
  S=\int d^4 x \mathcal{N}\left(-3a(\mathcal{N}^{-2}\dot{a}^2-k)+\frac{1}{2}a^3\mathcal{N}^{-2}\dot{\varphi}^2-a^3V(\varphi)\right),
\end{equation}
where $\mathcal{N}$ is the lapse function, and $k=0, \pm 1$ represents the
spacial curvature, and dot denotes the derivative with respect to
time. For simplicity, we have set $M_p^2\equiv 1/(8\pi G)=1$.

By varying the action with respect to the lapse function $\mathcal{N}$,
we obtain the Friedmann equation
\begin{equation}
  \label{eq:Friedmann}
  3H^2=\frac{1}{2}\dot{\varphi}^2+V(\varphi)-\frac{3k}{a^2},
\end{equation}
where after the variation, we have set $\mathcal{N}=1$. Varying the action with respect to $\varphi$ leads to the
scalar field equation of motion
\begin{equation}
  \label{eq:varphi}
  \ddot{\varphi}+3H\dot{\varphi}+V_{\varphi}=0,
\end{equation}
where the subscript $\varphi$ in $V$ denotes derivative with respect
to $\varphi$.

From the time derivative of (\ref{eq:Friedmann}) and using
(\ref{eq:varphi}), we get
\begin{equation}
  \label{eq:Hdot}
  \dot{H}=-\frac{1}{2}\dot{\varphi}^2+\frac{k}{a^2}.
\end{equation}

To construct the history space, we need to slice the $\mathcal{H}=0$
hypersurface of the phase space. A good way to do this is to choose
a constant $H$ surface  $H=H_S$ as a slicing \cite{GT06}, where $H_S$
is chosen low enough that it just above the end of inflation and the
universe evolves adiabatically from then on. To choose a constant $H$ slice is because
for the flat or open universe, and non-negative potential
$V(\varphi)$, each history trajectory crosses a constant $H$ surface
exactly once. And the reason for choosing $H$ low enough is that only this choice can result in a cutoff independent measure of
e-folds, and this choice is in agreement with the anthropic ``top
down'' approach to cosmology \cite{HH06}.

On a constant $H$ surface, the measure for the history
space takes the form
\begin{equation}
  \int_{H_S} \Omega_M \sim \int_{H_S} dp_{\varphi} \wedge d\varphi,
\end{equation}
where $p_{\varphi}\equiv a^3 \dot{\varphi}$ denotes the canonical
momentum for $\varphi$. It can be calculated that
\begin{equation}
  \label{eq:intOmega}
  \int_{H_S} \Omega_M \sim \int_{H_S} d\varphi da ~ 3a^2 \frac{6H_S^2-2V+4ka^{-2}}{\sqrt{6H_S^2-2V+6ka^{-2}}}.
\end{equation}

A divergence occurs in the large $a$ limit of
(\ref{eq:intOmega}). This is the infinity discovered in
\cite{HP88}. Following \cite{GT06}, we set a cutoff for the spacial
curvature to critical density ratio $\Omega_k\equiv -k/(a^2H_S^2)$
as
\begin{equation}
  \left|\Omega_k\right| \geq \Delta \Omega_k,
\end{equation}
The cutoff makes sense physically because a small enough $\Omega_k$ is
neither geometrically meaningful nor physically observable. As we are
working on a constant $H_S$ surface, the cutoff can be translated into
the cutoff of the scale factor
\begin{equation}
  a^2 \leq a_{\rm{max}}^2\equiv\frac{1}{\Delta \Omega_k H_S^2}.
\end{equation}
Recall that $\Omega_M=dA$, The measure can be reduced to a surface
integral around a constant $a$ surface of the constant $H_S$ history
space,
\begin{equation}
  \int_{H_S} \Omega_M=\int_{\partial H_S} A=\int_{\partial H_S}p_{\varphi}d\varphi=a_{\rm{max}}^3\int_{\partial H_S}\dot{\varphi}d\varphi.
\end{equation}

To investigate the probability distribution for inflation, now concentrate on an
history space volume element $A\sim \dot\varphi \Delta \varphi$. Where we have dropped the
$a_{\rm{max}}^3$ term as it is a constant. Since the variation
operation $\Delta$ is taken on a constant $H$ surface, it is
convenient to convert the time derivative $\partial_t$ to the
derivative with respect to the Hubble constant $\partial_H$, using
\begin{equation}
  \partial_t=\dot{H}\partial_H=-\frac{1}{2}\dot{\varphi}^2 \partial_H.
\end{equation}
Then we can take the advantage that $\partial_H \cdot \Delta=\Delta
\cdot \partial_H$.

Note that $H$ do not change when we move on the history space. Then
the equation (\ref{eq:Friedmann}) leads to a constraint for the history
space variation
\begin{equation}
  \dot{\varphi}\Delta\dot{\varphi}+V_{\varphi}\Delta\varphi=0,
\end{equation}

Given this constraint, the Hubble evolution for $A$ can be calculated as
\begin{equation}
  \label{eq:A1}
  \partial_H A =-\frac{3H}{\dot{H}} A.
\end{equation}
where we have neglected the spatial curvature energy density, because
it have to be small during the last e-folds of inflation.

Note that for the e-folding number $N$, $-H=\partial_t
N=\dot{H}\partial_H N$, the equation (\ref{eq:A1}) takes the form
\begin{equation}
  \partial_H A=3 A \partial_H N,
\end{equation}
which can be integrated out to give
\begin{equation}
  A= e^{3N} A(H_S).
\end{equation}
The above equation tells us that as we stand at the end of inflation
and track backwards with time, a volume in the history space expands
exponentially. In order not to break the slow roll condition along the
whole 60 e-folds' inflation, The volume element $A$ must lie in a exponentially narrow
corner in the constant $H_S$ history space. So the probability for
inflation is suppressed by the $\exp(-3N)$ factor. This suppression
shows that inflation is not as natural as we intuitively
think. It may have not solved the naturalness problems of the hot big bang
cosmology because of its unnatural nature, or there remains some unknown mechanism to
produce a exponentially sharp peak for the possibility distribution of
the history space.

\section{Generalized Single Field Models}

In this section, we consider the action
\begin{equation}
  \label{eq:S3}
  S=\int d^4 x
  \mathcal{N}\left\{-3a\mathcal{N}^{-2}\dot{a}^2+a^3f\left(\varphi,
    \mathcal{N}^{-1}\dot{\varphi}\right)\right\}.
\end{equation}
A good many inflation models can be described using this action. For
example, K-Inflation \cite{PDM99}, Phantom Inflation \cite{Phantom},
Inflation driven by the brane DBI action \cite{DBII, DBIII}, etc.

Choosing $\varphi$ as a canonical coordinate and using the proper time, the canonical momentum
for $\varphi$ takes the form

\begin{equation}
  p=\pi a^3, ~~~{\rm{where}}~ \pi\equiv f_{\dot{\varphi}}.
\end{equation}

Take variation with respect to $\mathcal{N}$ and $\varphi$, one obtains
\begin{equation}
  \label{eq:F3}
  3H^2=\dot{\varphi}\pi-f,~~~  \dot{\pi}+3H\pi-f_{\varphi}=0,~~~  \dot{H}=-\frac{1}{2}\dot{\varphi}\pi.
\end{equation}

Using (\ref{eq:F3}), the constraint for the variation in the history
space can be written as
\begin{equation}
\dot{\varphi}\Delta\pi=f_{\varphi}\Delta\varphi.
\end{equation}
And using the definition of $\pi$, we have the variation relation
\begin{equation}
  \Delta \dot{\varphi} = f_{\dot\varphi\dot{\varphi}}^{-1} (\Delta\pi-f_{\dot{\varphi}\varphi}\Delta\varphi).
\end{equation}
where we have assumed that $f_{\dot{\varphi}\dot{\varphi}}\neq 0$, in
order that $\varphi$ can be treated as a dynamical degree of freedom.

Now we take the cutoff as discussed in the last section, and reduce the
integration of the history space to the boundary integration
\begin{equation}
  \int_{H_S} \Omega_M=\int_{\partial H_S} A=\int_{\partial
    H_S}pd\varphi=a_{\rm{max}}^3\int_{\partial H_S}\pi d\varphi.
\end{equation}

Then it can be calculated that the variation of the volume element in
the history space $A\sim \pi
\Delta \varphi$ evolves along the constant $H$ surfaces as
\begin{equation}
  \partial_H A=-\frac{3H}{\dot{H}}A=3A \partial_H N
\end{equation}
So the conclusion is exactly the same as that of the last section.
In order to get $N$ e-folds' slow roll inflation, the volume
element in the history space should be exponentially fine turned.

It should be noticed that in this general case, there is the possibility
that even the history evolution is not slow rolling, accelerated
expansion with a large e-folding number can be achieved in models such
as the Kflation or the phantom inflation. But it is difficult to get a
scale invariant perturbation spectrum if the slow roll condition is
not satisfied \cite{Phantom}.

\section{Multi-Field Inflation Models}

Multi-field inflation models take an important part in the
inflationary model building.  In string theory, there can
be a number of scalar fields at the inflation scale. Phenomenally, in
multi-field models, slow roll condition is less stringent and can be satisfied
in more models \cite{PCZZ02}. Moreover, there are interesting inflation
models, like the hybrid inflation model \cite{Hybrid}, which requires
essentially more than one field. So it is useful to study the measure
for multi-field inflation and investigate the corresponding
probability.

The action for the multi-field inflation takes the form
\begin{equation}
  \label{eq:S2}
  S=\int d^4 x \mathcal{N}\left(-3a\mathcal{N}^{-2}\dot{a}^2+\frac{1}{2}a^3\mathcal{N}^{-2}\dot{\varphi_i}^2-a^3V(\varphi_i)\right),
\end{equation}
where the duplicate index $i$ is summed over the $n$ scalar fields.

Choosing to use the proper time, the canonical momentum for $\varphi_i$ is $p_i\equiv a^3
\dot\varphi_i$. And the equations of motion takes the form
\begin{equation}
  \label{eq:eom4}
  3H^2=\frac{1}{2}\dot\varphi_i^2+V,~~~\dot
  H=\frac{1}{2}\dot\varphi_i^2, ~~~\ddot\varphi_i+3H\dot\varphi_i+V_{\varphi_i}=0.
\end{equation}
The constraint for constant $H_S$ variation is
\begin{equation}
  \label{eq:C4}
  \dot\varphi_i\Delta\dot\varphi_i+V_{\varphi_i}\Delta\varphi_i=0.
\end{equation}
It can be checked by direct calculation that $\partial_H \cdot \Delta=\Delta
\cdot \partial_H$ is also true operating on $\dot\varphi_i$.
In this multi-field case
\begin{equation}
  A\sim\dot\varphi_1~\Delta\varphi_1\wedge\Delta\dot\varphi_2\wedge\Delta\varphi_2\wedge\ldots\wedge\Delta\dot\varphi_n\wedge\Delta\varphi_n.
\end{equation}

Using the constraint (\ref{eq:C4}), each term in $\partial_H A$ is
proportional to $A$, and $\partial_H A$ turns out to be
\begin{equation}
  \label{eq:pHA4}
  \partial_H
  A=3n\left\{1-\frac{1}{3n}\left(2\frac{\ddot\varphi_1}{H\dot\varphi_1}-\frac{\ddot H}{H\dot H}\right)\right\}A \partial_H N.
\end{equation}
In a multi-field inflation model, $\ddot H/(H \dot H)$ should also be
small and rolling slowly as in the single field case. If one assumes that
$\ddot\varphi_1/(H\dot\varphi_1)$ is also small and slow rolling, then the
integration can be carried out as
\begin{equation}
  A = e^{3nN} A(H_S),
\end{equation}
which shows that the departure from slow-roll evolves much faster
than that in the single field case. As a result, multi-field inflation
is much more unnatural then the single-field inflation with a much
smaller measure in the history space. This result is not surprising.
It is because from the first equation in (\ref{eq:eom4}), the Hubble
constant has contribution from the energy density of all inflation fields.
While from the third equation in (\ref{eq:eom4}), the Hubble constant appears as a friction in the evolution of
each single inflaton field. So in the multi-field inflation case, the
friction of each single field is contributed by all the fields, and
the history space for slow roll inflation is much more concentrated then the
single field models.

As analyzed in \cite{BHK03}, $\left|\ddot\varphi_1/(H\dot\varphi_1)\right| \ll 1$ may break
down in some multi-field inflation models. Now let's see whether a
fast rolling $\dot\varphi_1$ can result in something more natural.

If we want $\ddot\varphi_1/(H\dot\varphi_1)$ to cancel the exponential
expansion of the history space volume, we need
\begin{equation}
  \frac{\ddot\varphi_1}{\dot\varphi_1} \geq \frac{3n}{2}H,
\end{equation}
which amounts to demanding that $\left|\dot\varphi_1/a^{3n/2}\right|$
increases with time. As $n\geq 2$, $|\dot\varphi_1|$ must be increasing
faster than $a^3$ to make this cancellation possible. And this
cancellation need to be valid along the whole 60 e-folds of
inflation. It seems impossible for $\dot\varphi_1$ to behave like
this. So even a fast rolling $\dot\varphi_1$ can not make the
situation more natural.

A few words are in order here. We have picked a specific field $\varphi_1$
out of many other fields in studying the measure, this is just the result
of integrating out $p_{\varphi_1}$, namely, we have allowed $\dot\varphi_1$
to vary as much as possible. We could have picked out another field, then we
would be discussing the differential measure in a different region on the history space.

Now we see that the multi-field inflation is even more impossible than
the single field inflation. Then, if for anthropic principle or some other
reasons that a 60 e-folds' inflation has to have happened in our
history, it should be single field inflation rather than multi-field
inflation, because the latter has much smaller measure.

\section{Generalized Multi-Field Models}

In this section, we do the generalizations one step further to  consider the action
\begin{equation}
  \label{eq:S5}
  S=\int d^4 x
  \mathcal{N}\left\{-3a\mathcal{N}^{-2}\dot{a}^2+a^3f\left(\varphi_i,
    \mathcal{N}^{-1}\dot{\varphi_i}\right)\right\},
\end{equation}
which has the features of the actions in both Section 3 and Section 4.

Using the proper time, the canonical momentum for $\varphi$ takes the form
\begin{equation}
  p_i=\pi_i a^3, ~~~{\rm{where}}~ \pi_i\equiv f_{\dot{\varphi}_i},
\end{equation}
with the equations of motion
\begin{equation}
  \label{eq:F5}
  3H^2=\dot{\varphi}_i\pi_i-f,~~~  \dot{\pi}_i+3H\pi_i-f_{\varphi_i}=0,~~~  \dot{H}=-\frac{1}{2}\dot{\varphi}_i\pi_i.
\end{equation}
and the constraint for the history space variation
\begin{equation}
\dot{\varphi}_i\Delta\pi_i=f_{\varphi_i}\Delta\varphi_i.
\end{equation}
We assume that the matrix $f_{\dot\varphi_i \dot\varphi_j}$ has
inverse matrix. This should be true when all the constraints in
(\ref{eq:S5}) are solved and $\varphi_i$ only denotes the dynamical
degree of freedom. We use $f^{\dot\varphi_i \dot\varphi_j}$ as the inverse matrix of $f_{\dot\varphi_i
  \dot\varphi_j}$ Then it can be shown that
\begin{equation}
  \Delta \dot{\varphi}_i = f^{\dot\varphi_i \dot\varphi_j} (\Delta\pi_j-f_{\dot{\varphi}_j\varphi_k}\Delta\varphi_k).
\end{equation}
In this generalized case,
\begin{equation}
  A\sim\pi_1~\Delta\varphi_1\wedge\Delta\pi_2\wedge\Delta\varphi_2\wedge\ldots\wedge\Delta\pi_n\wedge\Delta\varphi_n.
\end{equation}
using the same technique developed in Section 3 and Section 4, one
finds
\begin{eqnarray}
\label{eq:GMA}
  \partial_H A&=&\left\{-\frac{3nH}{\dot H}+\frac{\dot\pi_1}{\dot H\pi_1}
    +\frac{f^{\dot\varphi_1
        \dot\varphi_i}\dot\pi_i}{\dot\varphi_1\dot H} +\frac{\dot\varphi_i\dot\pi_i}{2\dot
      H^2} + \frac{\pi_i f^{\dot\varphi_i\dot\varphi_k
       } \dot\pi_k}{2\dot H^2}
\right.\nonumber\\ & &\left.
-\frac{f^{\dot\varphi_1\dot\varphi_j}f_{\dot\varphi_j\varphi_k}\dot\varphi_k}{\dot
      H\dot\varphi_1}- \frac{\pi_i f^{\dot\varphi_i\dot\varphi_j}f_{\dot\varphi_j\varphi_k}\dot\varphi_k}{2\dot H^2} \right\}A
\end{eqnarray}
To see the implications of this equation, let us concentrate on the
double field inflation models. It is because it seems more difficult
to cancel the $-\frac{3nH}{\dot H}$ term for lager $n$. Lagrangian
densities like
$f=g(\varphi_1)\dot\varphi_1^2+h(\varphi_2)\dot\varphi_2^2$ are not of
special interest here, because they can be transformed into the case
discussed in Section 4 by a field redefinition. As another example,
let us consider the Lagrangian density
\begin{equation}
f=f\left(\varphi_1,\varphi_2,\frac{1}{2}(\varphi_1^2+\varphi_2^2)\right),
\end{equation}
in this case, the equation (\ref{eq:GMA}) takes the form
\begin{equation}
  \partial_H
  A=-6A\left\{1-\frac{1}{6}\left(2\frac{\ddot\varphi_1}{H\dot\varphi_1}+\frac{\dot f'}{Hf'}-\frac{\ddot H}{H\dot H}\right)\right\}\partial_H N ,
\end{equation}
where
\begin{equation}
f'\equiv \frac{\partial f}{\partial \left[\frac{1}{2}(\dot\varphi_1^2+\dot\varphi_2^2)\right]}.
\end{equation}
So a fast rolling $f'\dot\varphi_1^2$ is required to cancel the
exponential expansion of the volume of the history space.

To see the physical implications for this condition, consider
the DBI inflation model by \cite{DBIII}. The action of the DBI
inflation model is given by
\begin{equation}
  \label{eq:DBI}
  S_{\rm DBI}=-\int d^4x \left\{\tilde f^{-1}\left(\sqrt{1-\tilde f \dot\varphi^2}-1\right)+V(\varphi)\right\}
\end{equation}
where the angular motion of $\varphi_i$ has been ignored, so $\dot\varphi^2\equiv \dot\varphi_i^2$. Then $f'=1/\sqrt{1-\tilde
  f \dot\varphi^2}\equiv\gamma$ is just the relativistic factor
defined in \cite{DBIII}. From the spectral index
\begin{equation}
  n_s-1=4\frac{\dot H}{H^2}-2\frac{\ddot H}{H\dot H}+2\frac{\ddot\varphi}{H\dot\varphi}-2\frac{\dot\gamma}{H\gamma},
\end{equation}
We conclude that $\gamma$ should not be a fast rolling quantity along
the whole history of observable inflation. Moreover, from the equation $\dot
H=-\gamma\dot\varphi^2/2$, we see again that $\gamma$ can not be large
for a long time during inflation. So the cancellation of the
$e^{-3nN}$ factor can not be obtained.

\section{Conclusion and Discussions}

In this paper, we have reviewed the measure problem in cosmology. We
calculated the measure and the probability for inflation in single and multi-field models with
generalized Lagrangian density. It is shown that the measure for the single field
inflation and the corresponding generalizations are suppressed by a
factor of $\exp(-3N)$. While the $n$-field and generalized
multi-field inflation models has a measure proportional to
$\exp(-3nN)$.

This work can be understood in another way. Taking apart the discussion
for the measure and the slow roll condition, other parts of this paper can be thought of as a
proof of the attractor behavior of various kinds of generalized
inflation models. On the one hand, it is a proof that the attractor
behavior is very common in inflationary models. While on the other
hand, to take the measure into consideration, we see that it is
far from obvious for an attractor to be a natural solution in
cosmology. And it is just this early time attractor combined with the
requirement of slow roll that puts
inflation into a highly unnatural situation.

We did not study explicitly the inflation models with non-minimal
coupling to gravity \cite{nonminimal}. But these models do not seem to bring large
correction for the suppression factor. It is because through conformal
transformation, these non-minimal
coupled models are generally equivalent with the corresponding minimal
coupled inflation models with the same number or one more inflation
fields. Another reason not to consider these models in this work is
that, as the energy scale commonly drops during inflation, near
the end of inflation, the non-minimal coupling effect may not be so
important.

There are also inflation models with extra components or special spacetime
properties. Examples of this kind are inflation with holographic dark
energy \cite{HDE, HDEdata} or in the non-commutative spacetime
\cite{non2, noncommutative, non3}. These models do
not seem to change the results much either. Because in the former
example, the holographic dark energy is diluted during inflation so do
not seem to cause large corrections near the end of inflation. In the
latter case, although the spectrum for perturbations is greatly
modified in the non-commutative spacetime, the isotropic and
homogeneous inflating background do not change much because it belongs
to a lower
energy scale. So the corrections to the probability can not be large.

As a closing remark, we noticed that some non-inflationary models
do not share the small measure problem. One example is the cyclic universe
model \cite{cyclic}. Although the cyclic model is controlled by gravity coupled with a
scalar field, it do not have slow roll behavior backwards in time in the
cycle we live. So the key observation that the exponentially expansion
of the phase space volume breaks the slow roll condition do not apply
in the cyclic model. Nevertheless, the number of cycles in the cyclic universe
must be finite \cite{cyclic}, so it remains to explain how all the cycles begin
in the first place.

\section*{Acknowledgments}
This work was supported by grants of NSFC. We thank Bin Chen, Yi-Fu
Cai, Chao-Jun Feng, and Yushu Song for discussion.

\end{document}